\documentclass[aps, pre, twocolumn, floatfix, showpacs, footinbib]{revtex4-1}

\usepackage{calc}
\usepackage{graphicx}
\usepackage{color}
\usepackage{amsmath, amssymb}
\usepackage{array}
\usepackage{hyperref}

\def\abs#1{\left \vert #1 \right \vert}
\def\ave#1{\overline{#1}}

\def\etal{\textit{et al.\ }}

\graphicspath{{FIGURES/}}
\DeclareGraphicsExtensions{.eps}

\begin{document}

\title{Quantum chaos in one dimension?}
\author{L{\'a}szl{\'o} \surname{Ujfalusi}}
\author{Imre \surname{Varga}}
\email[Contact: ]{varga@phy.bme.hu}
\affiliation{Elm{\'e}leti Fizika Tansz{\'e}k, Fizikai Int{\'e}zet, 
             Budapesti M{\H{u}}szaki {\'e}s Gazdas{\'a}gtudom{\'a}nyi Egyetem,
             H-1521 Budapest, Hungary}
\author{D{\'a}niel \surname{Schumayer}}
\affiliation{Department of Physics, University of Otago, 730 Cumberland Street, Dunedin 9016, New Zealand}
\date{\today}

\begin{abstract}
   In this work we investigate the inverse of the
   celebrated Bohigas-Giannoni-Schmit conjecture. Using two inversion
   methods we compute a one-dimensional potential whose lowest $N$
   eigenvalues obey random matrix statistics. Our numerical results
   indicate that in the asymptotic limit, $N\to\infty$, the solution is
   nowhere differentiable and most probably nowhere continuous. Thus
   such a counterexample does not exist.   
\end{abstract}


\maketitle

\section{Introduction
         \label{sec:introduction}
        }

One of the main milestones of quantum chaos is the celebrated
Bohigas-Giannoni-Schmit (BGS) conjecture \cite{Bohigas1984}. This
conjecture forms a link between the statistical properties of the
spectra of quantum systems with chaotic classical limit and random
matrix theory (RMT):
   ``\emph{Spectra of time-reversal-invariant systems whose classical
           analogs are $K$-systems show the same fluctuation properties
           as predicted by GOE \footnote{The BGS conjecture contains two
           concepts which deserve further explanation. 
           {\emph{K-systems:}} In the mid-1950s Kolmogorov introduced
           a new measure, the metric entropy or K-entropy, to classify
           chaotic systems. Systems with positive K-entropy, such as
           Arnold's famous cat-map, have finite phase space with
           exponentially separating orbits. These systems are called
           K-systems. {\emph{GOE:}}
           The central objects of random
           matrix theory are the special ensembles of matrices. These
           ensembles are characterized by the size, the symmetry of the
           matrices and the domain of the individual random elements of
           these matrices, e.g. GOE is an abbreviation for the Gaussian
           Orthogonal Ensemble consisting real symmetric matrices for
           which the correlated probability of the diagonal elements is
           invariant under orthogonal transformation and follow a
           multi-dimensional normal distribution. This ensemble may,
           for example, correspond to spinless systems, invariant under
           time reversal.}}.''
%
%

This conjecture establishes a connection between chaotic systems
and RMT, i.e. Chaos $\rightarrow$ RMT. A number of numerical evidences
corroborate this conjecture \cite{osszefoglalo}, but a complete
analytical proof is still missing. As pointed out in 
Ref.~\cite{HannayAlmeida1984} this universality is a consequence
of the properties of the long periodic orbits whose semiclassical limit 
is also universal. The intimate relation between RMT statistics and 
its breakdown over energy scales of $\mathcal{O}(h/T)$ 
where $h$ is Planck's constant and $T$ is the period of the 
shortest periodic orbit, was emphasized in 
Refs.~\cite{HannayAlmeida1984, Berry1985}. The first attempt 
to prove the BGS conjectures is due to Andreev 
\etal \cite{Andreev1996a, Andreev1996b} but later it turned out 
to be incomplete. More recently the link between chaotic systems 
and RMT has been investigated using an improved semiclassical technique 
\cite{Muller2004, Muller2005, Keating2007}. Finally we have 
to mention that certain chaotic models show Poisson-statistics 
\cite{kivetel} allowing for the possibility of exceptions, as well.

Hence it is indeed very difficult to understand the exact
relationship between quantum chaos and RMT. Certainly it
should be helpful to investigate the inverse BGS conjecture,
i.e. does the presence of GOE level statistics ever lead to
chaos in the classical limit? This would be the opposite
link: RMT$\xrightarrow{?}$Chaos. This question has been
raised first by Wu \etal \cite{Wu1990}. The authors' aim was
to find a counterexample, i.e. a one-dimensional quantum system
with GOE level statistics. However, a one-dimensional system
is always integrable guaranteed by Liouville's theorem
\cite{Arnold}. The choice of $d=1$ is merely because it is
easier to do computation in one spatial dimension as compared
to $d=2$ or $d=3$. In addition the average level separation
is of the order of $\mathcal{O}(h^d)$ which allows a for 
semiclassical regime in $d>1$ only.
Despite the simplicity of $d=1$ the task is
not at all easy, finding a potential with a given spectrum leads
to an inverse Schr{\"o}dinger problem. There is almost no chance
to achieve this analytically, therefore Wu \etal used a numerical
method to find a one-dimensional potential with the first $N$
energy levels being predefined. Their process gave accurate
results for the case of GOE level statistics, hence they concluded
that there is an integrable system with GOE level statistics,
and provided a counterexample to the inverse BGS conjecture. 
Wu and Sprung later carried out a similar calculation~\cite{Wu93}
for a particular, asymptotically GUE spectrum represented by the
non-trivial zeros of the Riemann zeta function 
(see also~\cite{Ramani1995, Wu95, Brandon03, Schumayer2008, 
Schumayer2011}).
Their findings led to the concept of a fractal potential, which
may produce a chaotic spectrum.

Contrary to Wu \etal we can only state that the first $N$
eigenvalues of the potential obtained numerically show random
matrix type fluctuations leaving any other type of behavior for
the remaining infinite number of eigenvalues. Therefore we
re-investigate this problem more carefully with the emphasis
on the $N \rightarrow \infty$, limit. 

We examine certain statistical properties of the resulting potential
and conclude that no classical system can be associated with 
the constructed quantum potential. Our analysis not only confirms
a kind of fractal nature of the potential, already conjectured by
Wu and Sprung~\cite{Wu1990, Wu95}, but exceeds it by showing that 
the potential asymptotically loses its continuity and differentiability
and resembles the sum of randomly positioned Dirac-delta functions.

The structure of our article reads as follows. In Section
\ref{sec:NumericalMethods} we describe the numerical methods applied,
Section \ref{sec:results} compares the results provided by the
numerical methods. In Section \ref{sec:properties} we analyse the
$N \to \infty$ limit and Section \ref{sec:conclusion} is left for
conclusions.

\section{The numerical methods and \newline their properties
         \label{sec:NumericalMethods}
        }

We discuss the inverse BGS conjecture and start by calculating
all eigenvalues of a sufficiently large, real, symmetric matrix,
thereby creating a GOE spectrum. This raw spectrum has to be
unfolded in order to guarantee uniformity \footnote{Unfolding is
a standard procedure in RMT to scale the mean level spacing of
a given spectrum to be unity, thus different spectra become
comparable.}. As a result of this procedure, the average distance
between two neighboring levels in the unfolded spectrum is unity.
The first $N$ members of this unfolded spectrum are thereafter
denoted by $e_{1}$, $\cdots$, $e_{N}$ and remain unchanged.

We apply two techniques, the numerical method used also by Wu
\etal, and the dressing-transformation \cite{Ramani1995, Shabat1992,
Schumayer2008} in order to find a potential, whose first $N$ levels
are as close as possible to $e_{1}$, $\cdots$, $e_{N}$.
By changing $N$ we tried to conclude about the
$N \to \infty$ limit, since in that case the entire spectrum
should show GOE properties. Naturally, infinitely many potentials
may have the same first $N$ levels, nevertheless their shapes
coincide for the energy range determined by the prescribed
spectrum and differ only at higher energies.

Below we describe the numerical techniques; an optimisation and
an iterative one. These two approaches provide qualitatively and
quantitatively similar results concerning the most important
properties of the potential.

At first we implemented the same variational algorithm introduced
by Wu \etal \cite{Wu1990}. This method starts with an initial guess
$V_{0}(x)$ for the yet unknown potential $V(x)$ and one calculates
the energy eigenvalues of $V_{0}(x)$, which are denoted by
$\varepsilon_{1}$, $\cdots$, $\varepsilon_{N}$. This spectrum may
differ from the pre-defined energy levels, therefore in the next
step the potential is altered in a way, that its energy eigenvalues
fit the expected spectrum, $e_{1}$, $\cdots$, $e_{N}$, better. The
same procedure can be continued until convergence is reached.

This method requires an objective function, $G$, measuring the
distance between the actual and the pre-defined spectrum. One may
choose $G$ to be
\begin{equation} \label{eq:DefinitionOfG}
   G[V(x)] = \sum_{i=1}^{N}%
           { \left (
                      \varepsilon_{i} - e_{i}
             \right )^{2}
            }.
\end{equation}
$G$ implicitly depends on $V(x)$ through the set of eigenvalues,
$\varepsilon_{i}$, therefore the task is to minimize $G$
through the variation of $V(x)$. Moreover, $G$ is positive
semidefinite, and equals zero if and only if $\varepsilon_{i}
= e_{i}$ ($i=$1,2, \dots, $N$).

Employing a standard gradient method the potential has to be
updated as
\begin{equation} \label{potvalt}
   V_{\mathrm{new}}(x)
   =
   V_{\mathrm{old}}(x) -
   2 \,\eta \sum_{i=1}^{N}
                   {\left (
                             \varepsilon_{i} - e_{i}
                      \right )
                    \left \lbrack \Phi_{i}(x) \right \rbrack^{2}
                    }.
\end{equation}
If the initial potential, $V_{0}(x)$, is well chosen, the
algorithm converges and provides us with a potential, whose
first $N$ eigenvalues are very close to the pre-defined ones,
$\varepsilon_{i} \approx e_{i}$.

The average distance in the unfolded spectrum of a large GOE
matrix is unity, which resembles a harmonic oscillator,
therefore a parabolic form seems to be a good choice for
$V_{0}(x)$. 

Our other method is based on an interesting property of the
Schr{\"o}dinger equation: a bare free quantum system can be
``dressed up'' in an iterative manner, with a series of
potentials incorporating more and more bound states into the
system. Thereby this method is generally called {\emph{dressing
transformation}} \cite{Ramani1995, Shabat1992, Schumayer2008}.
Eventually this algorithm provides a potential, whose first $N$
lowest energy eigenstates are bound states with the pre-defined
energy levels, while all other states are scattering states,
i.e. $e_{1} < e_{2} < \cdots < e_{N} < 0$, and $e_{k}=0$ for
$k>N$. One may interpret this method differently, and say
that it shifts $N$ scattering states to negative energies,
converting them into bound states, while preserving the
original difference between the levels. It is important
to mention here, that theoretically this method is exact,
and only the implementation may introduce numerical error
into the final solution. 

In the remaining part of this section we comment on the
{\emph{uniqueness}} of the potential, on the {\emph{speed}}
of the algorithms, and finally on the {\emph{numerical
error}} accepted in our calculations.

{\emph{Uniqueness:}} If one intends to explore the interaction in a quantum
problem by analyzing scattering data, it is a crucial question whether the
experimental scattering data is compatible with many different potentials,
or just with one. For a generic non-relativistic, one-dimensional quantum
system Borg has rigorously proven that one spectrum (eigenvalues for the
bound states and reflection coefficient for the scattering states) is not
sufficient to uniquely reconstruct the corresponding potential \cite{Borg1946}.
However, from two spectra (obtained using two different boundary conditions)
one can determine the potential uniquely. Restricting ourselves
to even potentials only, $V(x) = V(-x)$, one spectrum becomes
sufficient to uniquely determine the potential \cite{Zakhariev1990}.
Unfortunately, noone can use an infinite set of spectrum in a numerical
algorithm, therefore uniqueness is lost apparently. Our physical intuition
suggests though, that if two potentials support the same bound states, then
these potentials ought to be identical at least around their bottoms.

This expectation is valid and justified numerically in our calculation.
The potential generated by the Wu-method and by the dressing transform
-- after an insignificant energy shift -- differ only at their ``semiclassical
edges'' which can be characterized by $x_{\mathrm{turn}}$, where
$V(x_{\mathrm{turn}}) = e_{N}$, i.e. where a classical particle with energy
$e_{N}$ would turn back in the given potential.

The difference between the potentials is expected outside the 
$[-x_{\mathrm{turn}}, x_{\mathrm{turn}}]$ interval, since the Wu-method should
reproduce the parabolic potential characteristic of the harmonic oscillator,
while in this form of the dressing transformation the first derivative of
the potential must approach zero asymptotically, therefore the potential
becomes constant.

{\emph{Speed:}} According to our experience the Wu method was very accurate;
the error decreases rapidly in each consecutive iteration. Despite its
precision, it becomes slow for increasing $N$ mainly due to the repeated
calculation of every eigenstates, $\Phi_{i}(x)$ ($i=$ 1, 2, \dots, $N$).
This method seems to be prohibitively impractical for $N$ being higher
than few hundreds. On the contrary the dressing transformation is very
fast, but for large $N$, its accuracy decreased considerably. 

{\emph{Numerical error:}} Both methods introduce a numerical error
due to the discretization. It is, therefore, necessary to check
whether the potential calculated reproduces the spectrum prescribed.
If the spectrum $\varepsilon_{i}$ does not match the original set of
eigenvalues, $e_{i}$, the appropriate parameters of the algorithms
must be adjusted: the variational step size, $\eta$, in case of the
Wu-method, and the discretization step-size of the Runge-Kutta method
for the dressing transformation.

We stopped the algorithms when the average error on each energy level
has been reduced below a fixed threshold, namely
\begin{equation} \label{eq:ConditionOnError}
   \frac{1}{N}
   \sum_{i=1}^{N}
       {\abs{e_{i}-\varepsilon_{i}}}
   < 10^{-5}.
\end{equation}
This seemingly acceptable order of numerical error is easily met by
the Wu-method, but represents a stringent condition for the dressing
transform, the later of which needed small step-size, $\sim 2 \times
10^{-4}$, for relatively high values of $N$. We have also investigated
the accumulation of error and have found that the sum in
Eq.~(\ref{eq:ConditionOnError}) collects the main contributions at
high energies, i.e. where the two methods are expected to deviate
from each other.

\section{Results
         \label{sec:results}
        }

\begin{figure}[hbt!]
   \includegraphics[width=\textwidth/2-4mm]{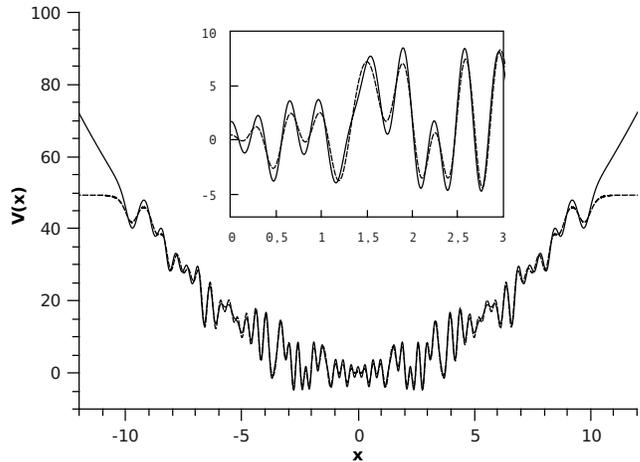}
   \caption{\label{wuschu50}
            The potentials calculated with the Wu-method (solid
            line) and the dressing transformation (dashed line)
            for $N=50$ are shown. The potentials provided by these
            methods are almost identical in the [-10, 10] range.
            Inset shows the same potentials in the [0, 3] interval.
           }
\end{figure}

This section provides the results. First let us take a look at
Fig.~\ref{wuschu50} which depicts the potentials obtained by the
Wu-method and the dressing transformation for $N=50$.

\begin{figure}[htb!]
   \includegraphics[width=\textwidth/2-4mm]{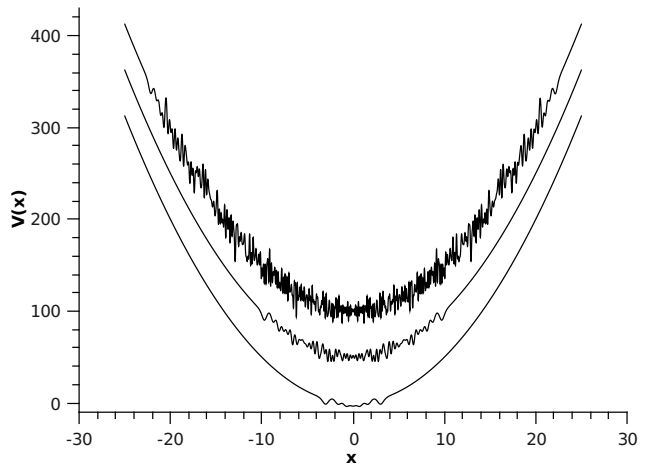}
   \caption{\label{pot550250}
            The potential calculated with the Wu method for $N=5$ (bottom),
            $50$ (middle) and $250$ (top). The graphs for $N=50$ and $N=250$
            are shifted vertically.
           }
\end{figure}

\begin{figure*}[tbh!]
   \includegraphics[width=\textwidth]{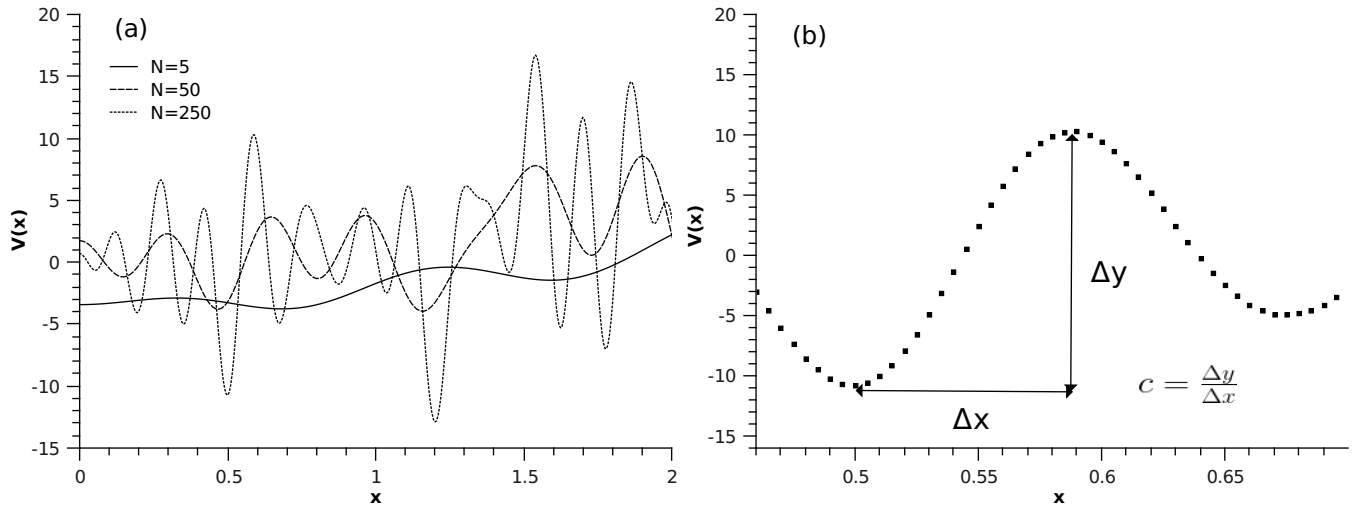}
   \caption{\label{potokdeltaxyc}
            (a) The potential calculated with the Wu method for $N=5$
            (continuous line), $N=50$ (dashed line), $N=250$ (points)
            in the interval $[0; 2]$.
            (b) The meaning of $\Delta x$ and $\Delta y$.
           }
\end{figure*}

\begin{figure}[tbh!]
   \includegraphics[width=\textwidth/2-4mm]{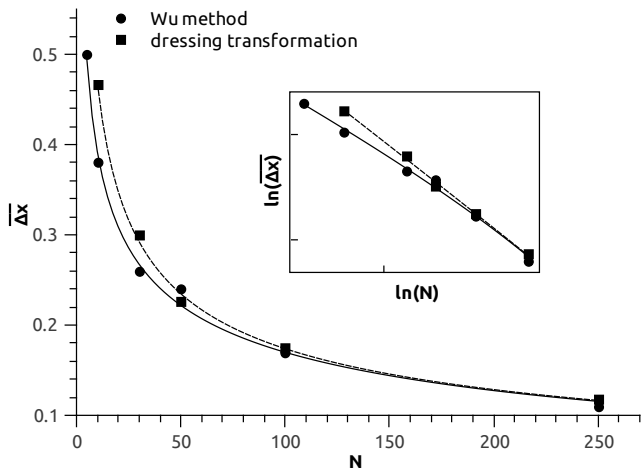}
   \caption{\label{AverageDeltax}
            The average distance of two adjacent extrema, $\ave{\Delta x}$,
            is plotted as a function of $N$ for both methods. The solid and dashed 
            lines are plotted only to guide the eye and are obtained by fitting 
            the function $A+BN^{-C}$ onto the data. 
            The inset is the same figure on a log-log scale.
           }
\end{figure}

\begin{figure}[tdh]
   \includegraphics[width=\textwidth/2-4mm]{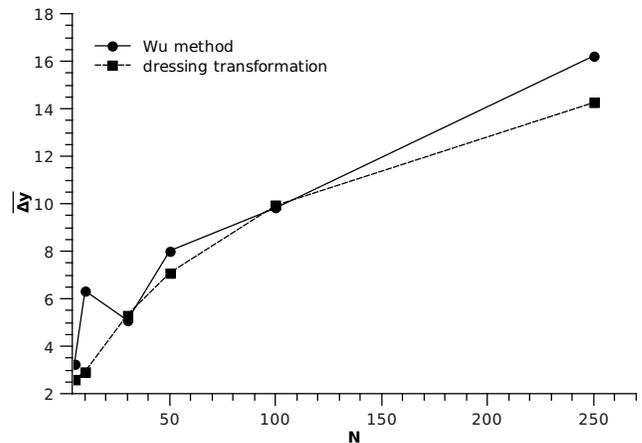}
   \caption{\label{AverageDeltay}
            The average vertical ``jump'' between two adjacent extrema, $\ave{\Delta y}$,
            is depicted as a function of $N$ obtained using both methods.
           }
\end{figure}

\begin{figure}[tb!]
   \includegraphics[width=\textwidth/2-4mm]{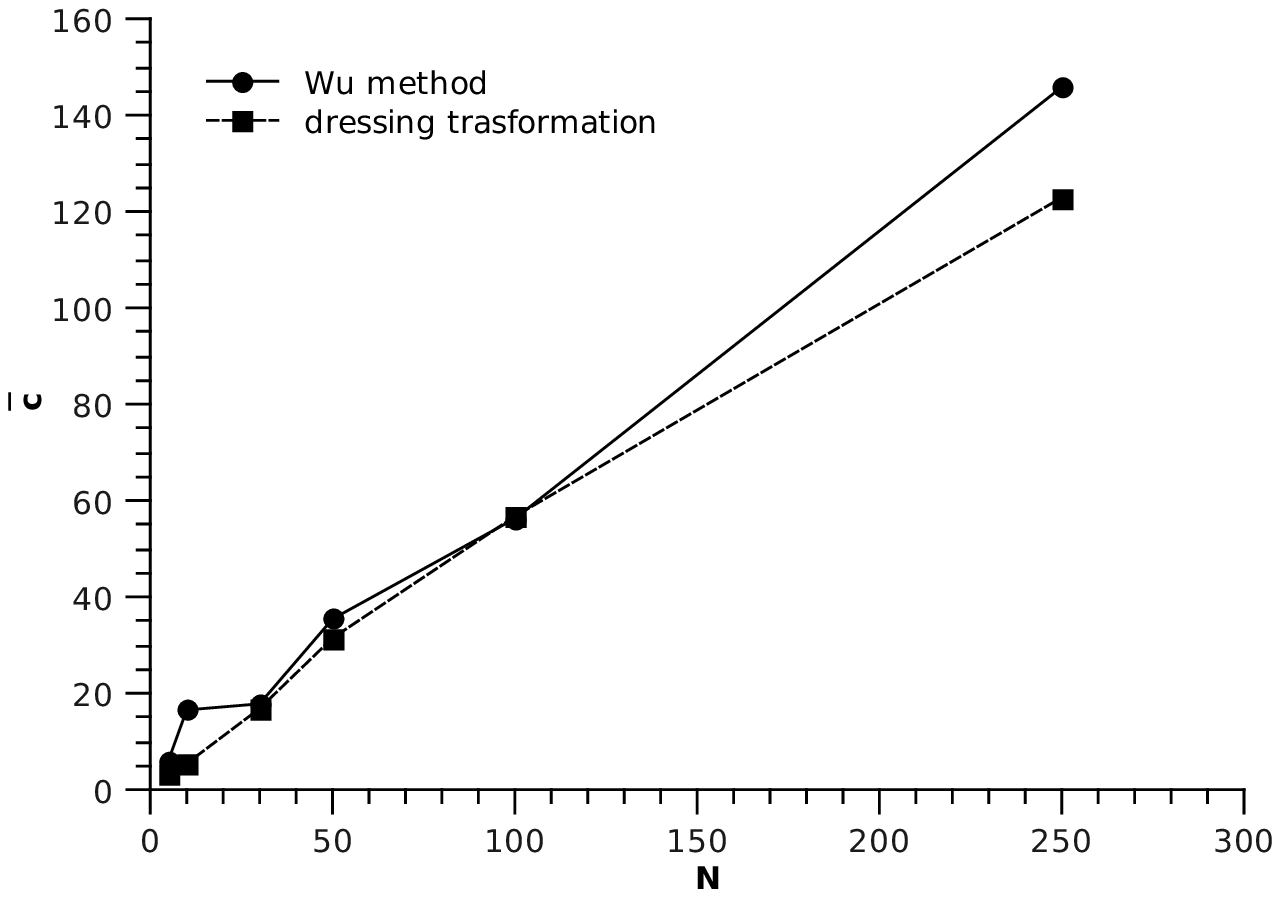}
   \caption{\label{E(c)}
            The average of $c$ as a function of $N$ with
            the Wu method and  the dressing transformation.
           }
\end{figure}

Note that the two methods do indeed give very similar, symmetric
potentials, as we expected based on Borg's theorem. The dressing
transformation provides a symmetric potential {\it a priori},
while the Wu method is started from a symmetric potential which
property is inherited in each iteration. More importantly, one may
separate the potential into two parts; a parabola and an oscillatory
component, the latter of which will be called fluctuation.
\begin{equation} \label{vfl}
   V(x) = \ave{V(x)} + V_{\mathrm{fl}}(x).
\end{equation}
Due to the GOE level fluctuations of the first $N$ eigenvalues the
parabola becomes oscillatory in a finite region that is connected
to the $N$th level, i.e. to the turning point $x_{\mathrm{turn}}$
for which $V(x_{\mathrm{turn}}) = E$. In quantum mechanics the wave
function approaches zero very rapidly for $\abs{x} > x_{\mathrm{turn}}$,
thus the oscillatory region is simply the interval of $[-x_{\mathrm{turn}};
x_{\mathrm{turn}}]$ if $E = \varepsilon_{N}$. Hence the average
potential becomes wavy in the region according to $x_{\mathrm{turn}}$
corresponding to the $N$-th level. By increasing $N$, the turning
point, $x_{\mathrm{turn}}$, increases too, therefore in the $N \to
\infty$ limit the whole potential is expected to be wildly oscillatory.
This expectation is supported by Fig.~\ref{pot550250}, although the
largest value of $N$ used is only 250. However, not only the width
of the oscillatory region increases, but the amplitude of the
fluctuation term seems to be larger and larger, see Figs.~\ref{pot550250}
and \ref{potokdeltaxyc}(a).
In the following we wish to classify the oscillatory component,
$V_{\mathrm{fl}}(x)$, by measuring the density of the fluctuations.
For this reason the distance between two consecutive extrema could 
serve as a good measure, see Fig.~\ref{potokdeltaxyc}(b), therefore,
we defined $\Delta x$ and $\Delta y$ as the distances between two
adjacent extrema in the $x$ and $y$ directions.
For a fixed value of $N$, the average value of $\Delta x$ and $\Delta y$
characterizes the fluctuation of $V(x)$. The average of $\Delta x$ is
plotted in Figure \ref{AverageDeltax} for both methods. The graphs
suggest a power-law dependence on $N$. The least-squares fit, using
$\ave{\Delta x} = A + B N^{-C}$ is also plotted to guide the eye.
Provided this dependence is suitable also in the asymptotic limit,
$N\to\infty$, the average width of oscillation in $V_{\mathrm{fl}}(x)$
approaches zero, since $A\approx 0$, while the average amplitude of 
fluctuation, $\ave{\Delta y}$, remains positive, see Fig.~\ref{AverageDeltay}. 
\begin{figure}[htb!]
   \includegraphics[width=\textwidth/2-4mm]{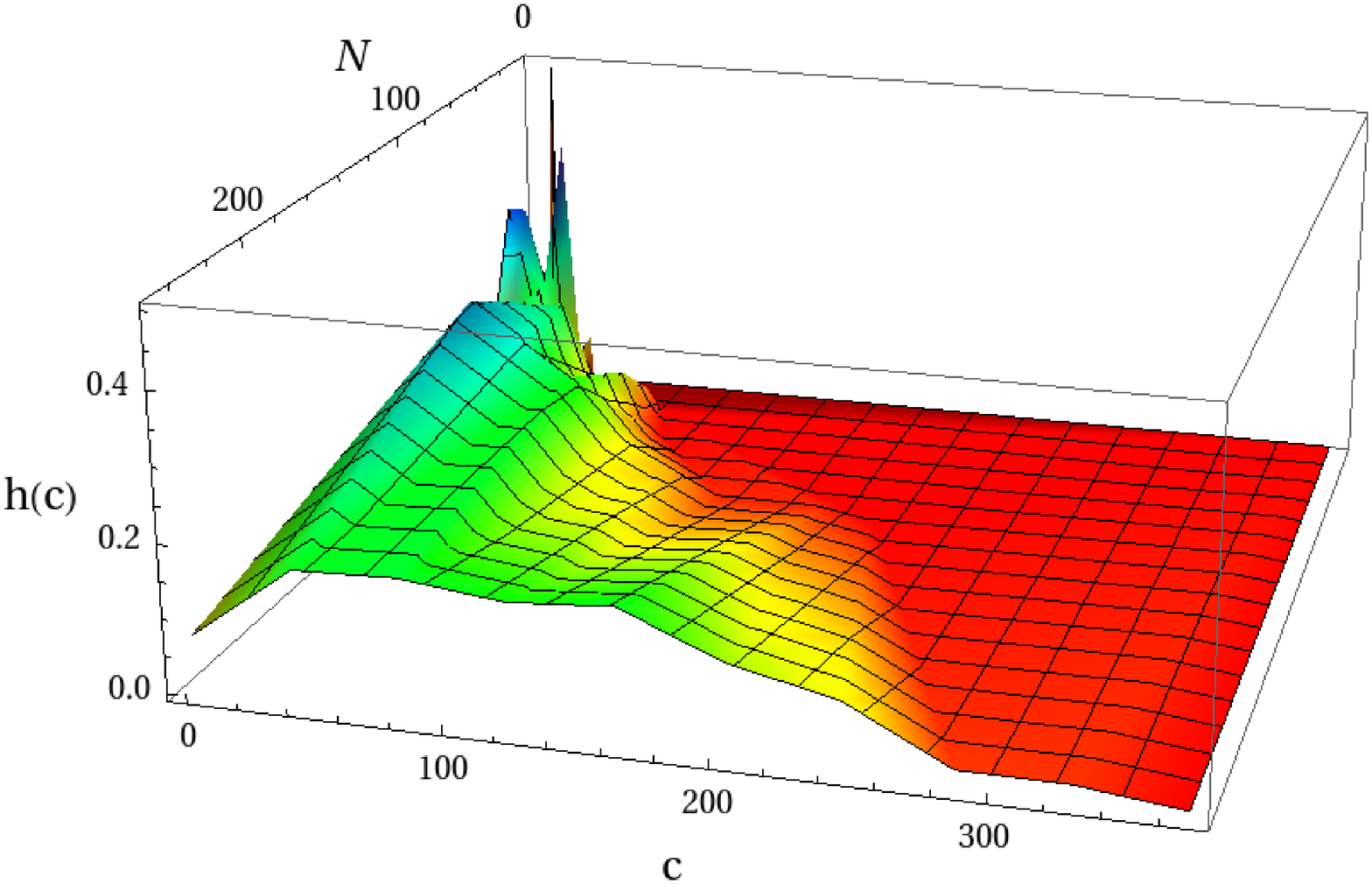}
   \caption{\label{derivalteloszlas}
            (Color online) The histogram of the parameter $c$ is depicted
            for different values of $N$.
           }
\end{figure}
This asymptotic behavior naturally poses the question: does the potential
remain differentiable as $N \to \infty$? In order to answer this question
we introduce a new quantity with the following definition: $c = \Delta y/
\Delta x$. This fraction, at any finite $N$, is not equal to the usual
difference fraction, because neither $\Delta x$, nor $\Delta y$ is allowed
to change continuously as in calculus, but here they have a finite and fixed
local value. For any potential calculated for a fixed $N$, we have a finite
set of $c$ values to analyze statistically. In Figure \ref{derivalteloszlas}
we show the histogram of $c$ for different values of $N$. One can see that
this histogram broadens as $N$ increases. Conclusively, $\ave{c}$, the
average of $c$, diverges proportional to $N$, which is seen in Fig.~\ref{E(c)}. 
The behavior of $\ave{\Delta x}$ and $\ave{c}$ described above
allows us to conclude: in the $N\to\infty$ limit the whole potential will
be rapidly oscillatory, because $x_{\mathrm{turn}}$ is growing with $N$. 
Considering any finite interval of $x$, $\ave{\Delta x} \to 0$ as $N \to
\infty$, therefore $c$ becomes the derivative of $V(x)$. From the numerical
observations mentioned above one can conclude ({\it i}) there are infinitely
many extrema, thus infinitely many values of $c$ in a finite interval of $x$,
and ({\it ii}) a finite part of the $c$-s, still infinitely many, diverge
with increasing $N$, since $\ave{c}$ diverges as well. These findings
suggest that the potential is nowhere differentiable. Such a potential cannot
be compatible with the Hamiltonian formalism of classical mechanics, especially
with the Hamilton-equation $\dot{p} = -{\partial H}/{\partial q}$. Therefore
Wu's counterexample for the inverse BGS conjecture is not acceptable, since
the classical limit of a potential whose {\emph{entire}} spectrum obeys GOE
level statistics makes no physical sense. Subsequently, no meaning can be
assigned to the question whether it leads to chaotic dynamics or not.

\section{Mathematical properties of the $N\to\infty$ limit
         \label{sec:properties}
        }

In the previous section we demonstrated that a quantum potential
whose energy eigenvalues follow GOE level statistics is nowhere
differentiable in the $N \to \infty$ limit. Can one classify the
behavior of this potential any further? Is there a characteristic
width of the oscillations appearing in the potential? If `yes',
what is the asymptotic value of this characteristic width as $N$
increases?

\begin{figure}[htb!]
   \includegraphics[width=\textwidth/2-4mm]{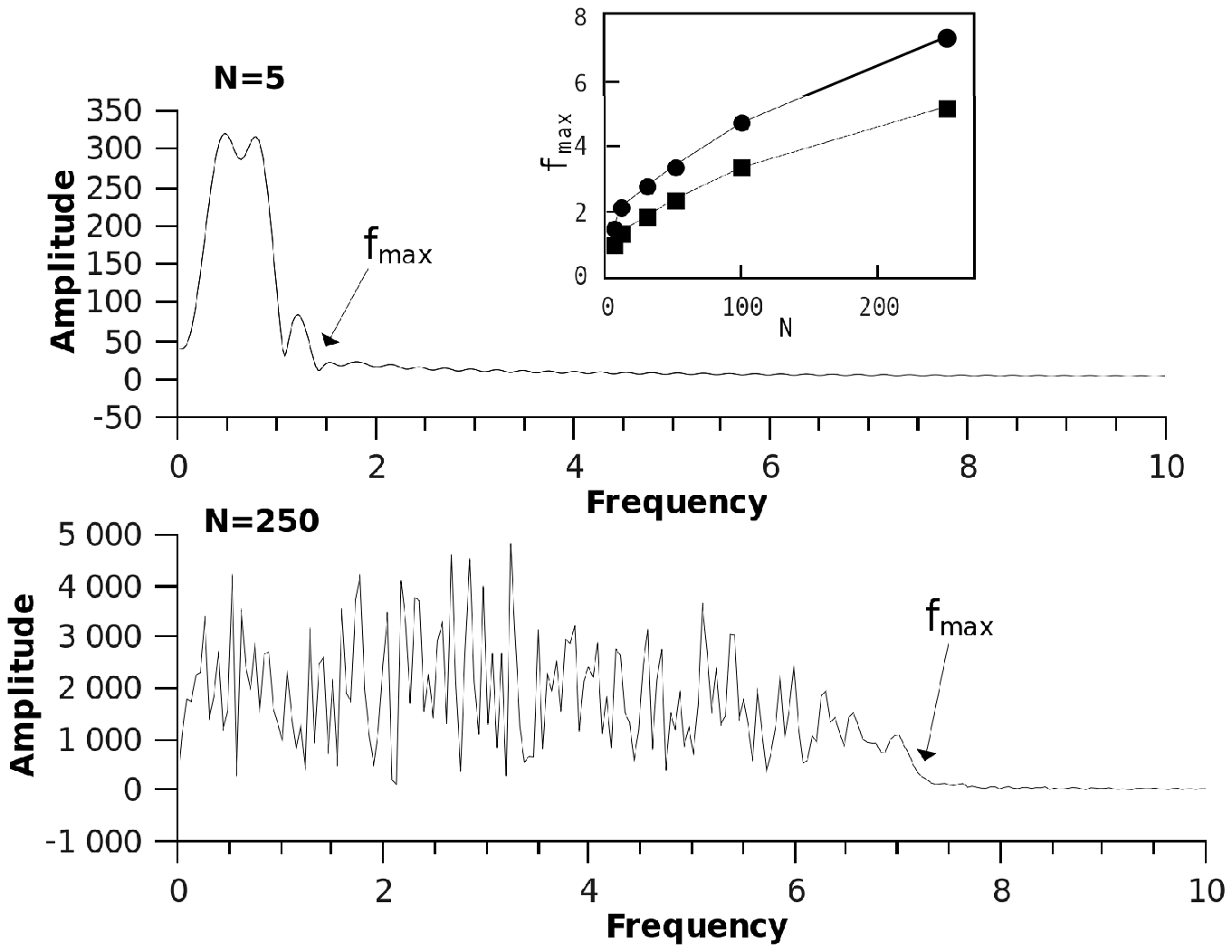}
   \caption{\label{fourier}
            The Fourier-transform of $V_{\mathrm{fl}}(x)$ at
            $N=5$ (upper graph) and at $N=250$ (lower graph).
            $\mathcal{F}[V_{\mathrm{fl}}(x)]$ gets wider as $N$
            increases. $\mathcal{F}[V_{\mathrm{fl}}(x)]$ vanishes
            beyond a characteristic wavenumber for every $N$ denoted
            as $f_{\mathrm{max}}$. Inset: $f_{max}$ as a function of
            $N$ with both methods. The two curves show very similar
            behavior.
           }
\end{figure}

\begin{figure}[tdh]
   \includegraphics[width=8cm]{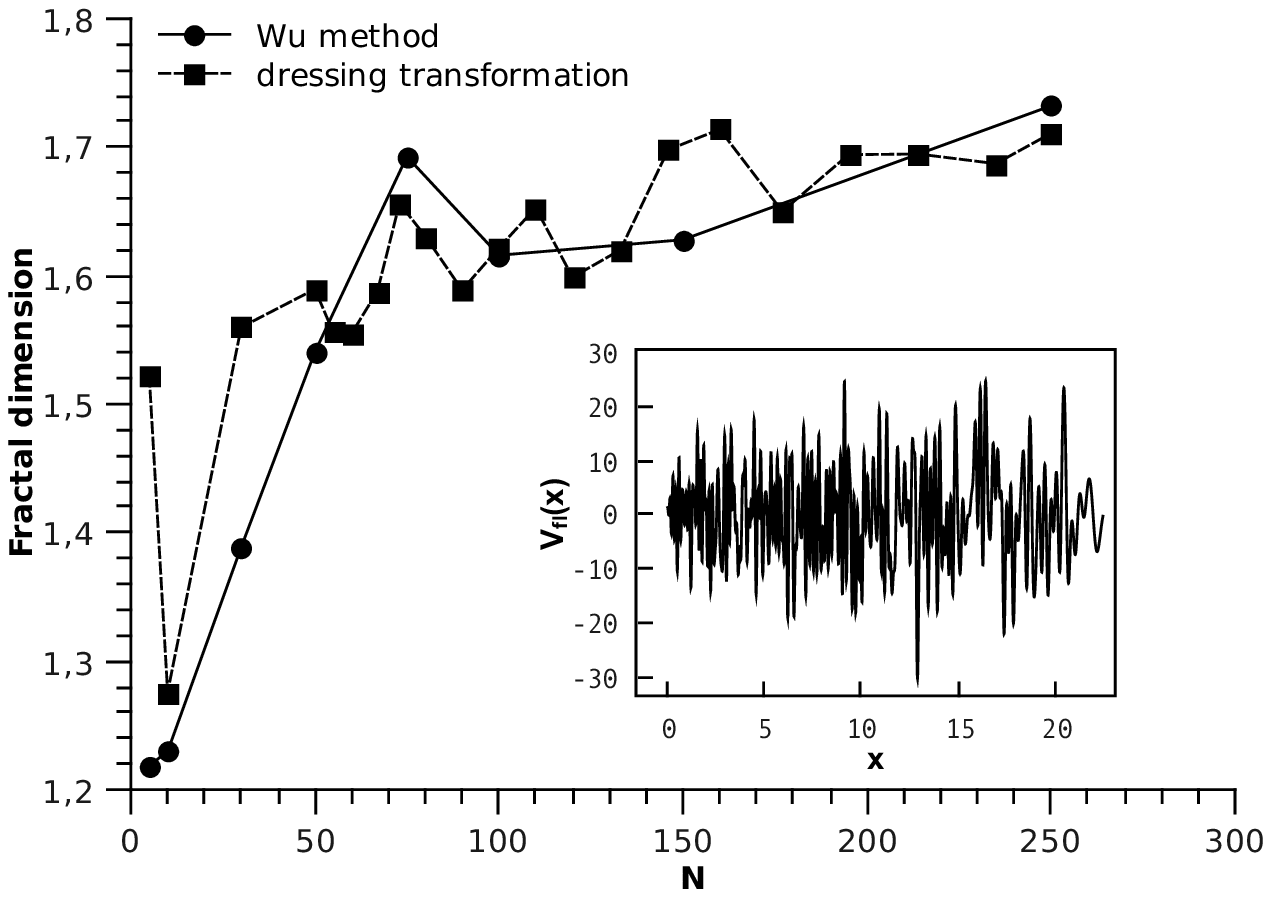}
   \caption{\label{fracdim}
            The box-dimension of $V_{\mathrm{fl}}(x)$ as a
            function of $N$. Inset: $V_{\mathrm{fl}}(x)$ at $N=250$.
           }
\end{figure}

\begin{figure}[t!]
   \includegraphics[width=\textwidth/2-4mm]{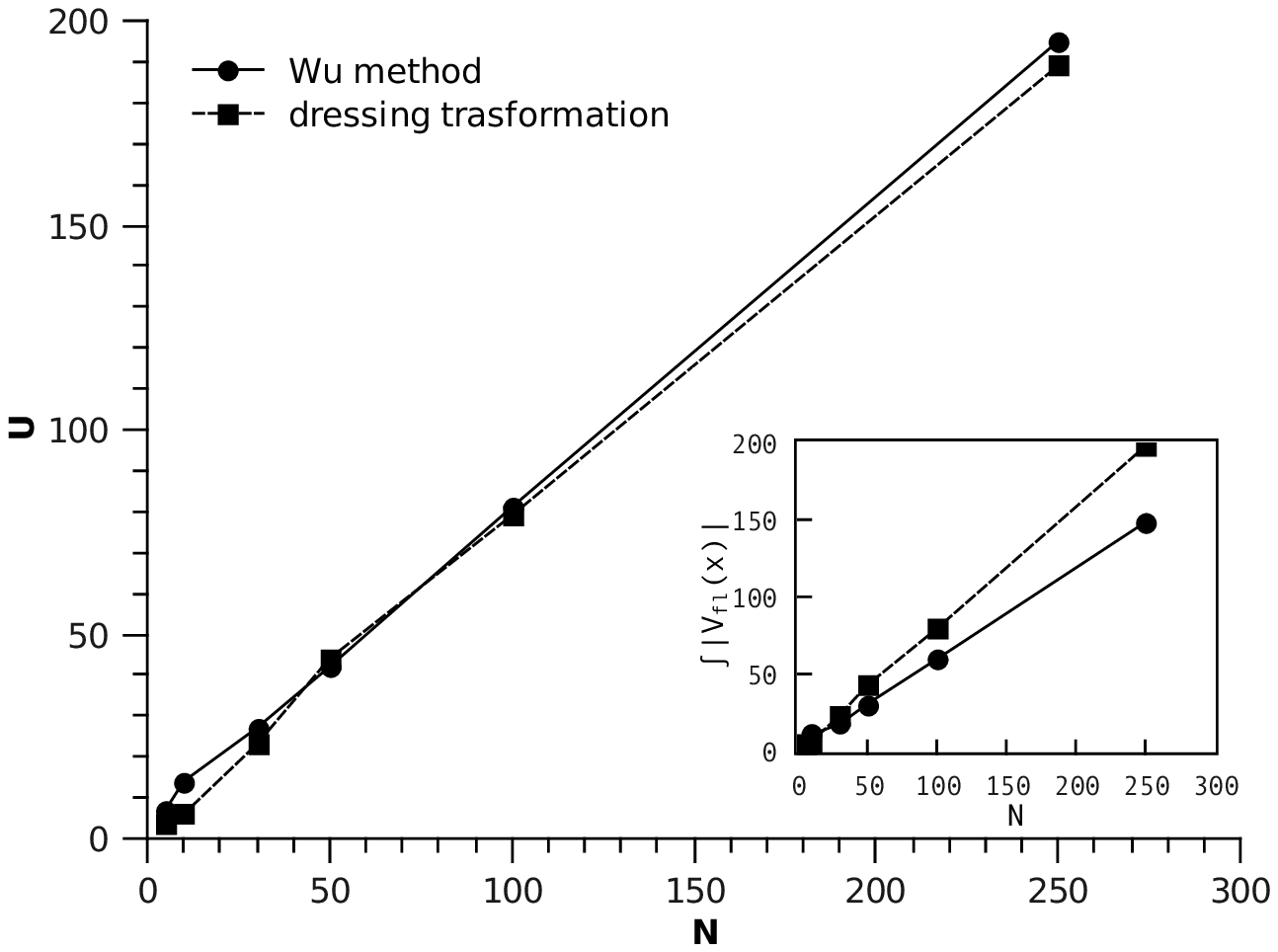}
   \caption{\label{wuschuintVfl-N}
            The number of nodes of the potential is plotted as a function
            of $N$. Inset shows the absolute value of $ V_{\mathrm{fl}}(x)$
            integrated over the entire $x$-axis. The solid and dashed lines
            correspond the same method in both graphs.
           }
\end{figure}

In order to answer these questions we are going to focus on the
fluctuation part of the potential and calculate for instance 
its numerical
Fourier transform, $\mathcal{F}[V_{\mathrm{fl}}(x)]$. The Fourier
components, for different values of $N$, are depicted in
Fig.~\ref{fourier}.
It is apparent that the Fourier-transform differs from zero only in
an interval [0, $f_{\mathrm{max}}$]. This upper bound can be thought
of as a characteristic frequency, which translates into a
characteristic width. As $N$ increases $f_{\mathrm{max}}$ increases
as well (see inset of Fig.~\ref{fourier}) therefore waves with smaller
and smaller wavelength appear in the graph of the potential. In the
$N \to \infty$ limit we expect that every wavelength will appear, so
the graph of the potential is going to lose its characteristic length
scale. This is exactly a particular property of fractals. In
Fig.~\ref{fracdim} the fractal dimension of the potential is plotted,
obtained by using the standard box-counting algorithm. Both methods
employed yield similar behavior: a slowly increasing box-dimension as
$N$ increases. Unfortunately the computed dimension fluctuates too much
to derive a definite conclusion. However, the dimension is most likely
not smaller than 1.6 in the $N \to \infty$ limit. This is comparable
to the value of 1.5 obtained in an earlier work for the potential
derived from the nontrivial zeros of the Riemann zeta-function~\cite{Wu93}.
This means that the potential strongly shades the complete
plane for large-enough $N$, which can be seen in the inset of
Fig.~\ref{fracdim}.
Earlier we concluded that our potential is nowhere differentiable.
Furthermore continuity would require $\ave{\Delta y}$ to decrease
and approach zero as the $N \to \infty$. In Fig.~\ref{AverageDeltay}
one can see that $\ave{\Delta y}$ does not tend to zero. If
$\ave{\Delta y}$ increases indefinitely, the graph of the potential
is not even bounded. In order to analyse this possibility, we also
calculated the area under the graph of the potential defined as the
definite integral of its absolute value $\int{\abs{V_{\mathrm{fl}}(x)}
dx}$. Fig.~\ref{wuschuintVfl-N} shows the number of nodes, $U$, while 
the inset depicts the total area under the graph of the potential
of the same graph. It is apparent from Fig.~\ref{wuschuintVfl-N} that
both the area and the number of nodes increases linearly with $N$. 
The area under {\emph{one}} oscillation is the fraction of the slopes
of these lines, which turns out to be a constant, irrespectively of the
method we employ. Since for increasing $N$ the average width of
oscillations tends to 0, $\ave{\Delta x} \to 0$, but the area under
{\emph{one}} oscillation remains constant, one might conclude that
the height of a wave ought to tend to infinity, i.e. $\ave{\Delta y}
\to \infty$, i.e. the potential has a Dirac-delta-like
singularity at every point in the $N \to \infty$ limit. In order to
confirm this conclusion, we have calculated the Fourier-transform of
sum of sinc-functions centered at random $x$ value. The result
of this calculation showed qualitatively similar behavior as the
one presented in Fig.~\ref{fourier}.

\section{Conclusions
         \label{sec:conclusion}
        }

In this paper we investigated the inverse BGS conjecture and the
validity of the counterexample given by Wu \etal The Hamiltonian of
our model takes its most standard form $p^{2}/2m + V(x)$, and we
created one-dimensional quantum potentials, $V(x)$, whose first
$N$ energy levels follow GOE properties. We investigated the $N
\to \infty$ limit and showed that such a potential is nowhere
differentiable and nowhere continuous, moreover may it has a
singularity at ``every'' point. No sensible physical meaning
can be associated with such an ill-behaving potential, therefore,
the question whether it is chaotic or not, cannot be interpreted
either. We mention here, however, the logically plausible, but
improbable scenario that a more complicated Hamiltonians,
e.g. Harper's model $\cos{(p)} + \cos{(x)}$ \cite{Harper1955,
Hofstadter1976}, may reproduce the GOE statistics. 

We used a method originally employed by Wu \etal, and the dressing
transformation in order to find a potential whose entire spectrum
obeys GOE statistics. These methods are based on different ideas,
nevertheless they provided quantitatively the same results for the
relevant measures investigated. The range of $N$ was between $5$
and $250$. 

We further analyzed the $N \to \infty$ limit by calculating the Fourier
spectrum of the fluctuating part of the potential. We demonstrated
that its Fourier components do not tend to zero at large magnitudes
of $f$. Therefore the potential loses its characteristic length scale,
and its structure resembles that of a fractal, the dimension is above
$1.6$. This finding also supports our conclusion, namely a potential
having GOE spectrum does not have a classical analogue, therefore it
is not an acceptable counterexample against the inverse BGS conjecture.

\begin{acknowledgments}
   The authors are indebted to Dr. P. B{\'a}lint and Prof. B. Eckhardt
   for fruitful and enlightening discussions. The authors thank one of 
   the referees for pointing out Refs. 
   \cite{HannayAlmeida1984, Berry1985, Keating2007}. 
   This work was partially
   financed by the Hungarian Research Fund (OTKA) grants K73361 and
   K75529, the New Hungary Development Plan (Project ID:
   T{\'A}MOP-4.2.1/B-09/1/KMR-2010-0002) and the Alexander von
   Humboldt Foundation. D. S. acknowledges the financial support 
   by the University of Otago.
\end{acknowledgments}

\bibliographystyle{apsrev4-1}
\bibliography{integrability1d}

\begin{thebibliography}{10}%
\makeatletter
\providecommand \@ifxundefined [1]{%
 \ifx #1\undefined \expandafter \@firstoftwo
 \else \expandafter \@secondoftwo
\fi
}%
\providecommand \@ifnum [1]{%
 \ifnum #1\expandafter \@firstoftwo
 \else \expandafter \@secondoftwo
\fi
}%
\providecommand \enquote [1]{``#1''}%
\providecommand \bibnamefont  [1]{#1}%
\providecommand \bibfnamefont [1]{#1}%
\providecommand \citenamefont [1]{#1}%
\providecommand\href[0]{\@sanitize\@href}%
\providecommand\@href[1]{\endgroup\@@startlink{#1}\endgroup\@@href}%
\providecommand\@@href[1]{#1\@@endlink}%
\providecommand \@sanitize [0]{\begingroup\catcode`\&12\catcode`\#12\relax}%
\@ifxundefined \pdfoutput {\@firstoftwo}{%
 \@ifnum{\z@=\pdfoutput}{\@firstoftwo}{\@secondoftwo}%
}{%
 \providecommand\@@startlink[1]{\leavevmode\special{html:<a href="#1">}}%
 \providecommand\@@endlink[0]{\special{html:</a>}}%
}{%
 \providecommand\@@startlink[1]{%
  \leavevmode
  \pdfstartlink
   attr{/Border[0 0 1 ]/H/I/C[0 1 1]}%
   user{/Subtype/Link/A<</Type/Action/S/URI/URI(#1)>>}%
  \relax
 }%
 \providecommand\@@endlink[0]{\pdfendlink}%
}%
\providecommand \url  [0]{\begingroup\@sanitize \@url }%
\providecommand \@url [1]{\endgroup\@href {#1}{\urlprefix}}%
\providecommand \urlprefix [0]{URL }%
\providecommand \Eprint[0]{\href }%
\@ifxundefined \urlstyle {%
  \providecommand \doi [1]{doi:\discretionary{}{}{}#1}%
}{%
  \providecommand \doi [0]{doi:\discretionary{}{}{}\begingroup
  \urlstyle{rm}\Url }%
}%
\providecommand \doibase [0]{http://dx.doi.org/}%
\providecommand \Doi[1]{\href{\doibase#1}}%
\providecommand \bibAnnote [3]{%
  \BibitemShut{#1}%
  \begin{quotation}\noindent
    \textsc{Key:}\ #2\\\textsc{Annotation:}\ #3%
  \end{quotation}%
}%
\providecommand \bibAnnoteFile [2]{%
  \IfFileExists{#2}{\bibAnnote {#1} {#2} {\input{#2}}}{}%
}%
\providecommand \typeout [0]{\immediate \write \m@ne }%
\providecommand \selectlanguage [0]{\@gobble}%
\providecommand \bibinfo [0]{\@secondoftwo}%
\providecommand \bibfield [0]{\@secondoftwo}%
\providecommand \translation [1]{[#1]}%
\providecommand \BibitemOpen[0]{}%
\providecommand \bibitemStop [0]{}%
\providecommand \bibitemNoStop [0]{.\EOS\space}%
\providecommand \EOS [0]{\spacefactor3000\relax}%
\providecommand \BibitemShut [1]{\csname bibitem#1\endcsname}%
\bibitem{Bohigas1984}%
  \BibitemOpen
  \bibfield{author}{%
  \bibinfo {author} {\bibfnamefont{O.}~\bibnamefont{Bohigas}}, \bibinfo
  {author} {\bibfnamefont{M.~J.}\ \bibnamefont{Giannoni}},\ and\ \bibinfo
  {author} {\bibfnamefont{C.}~\bibnamefont{Schmit}},\ }%
  \bibfield{journal}{%
  \Doi{10.1103/PhysRevLett.52.1}{\bibinfo {journal} {Phys. Rev. Lett.}}\ }%
  \textbf{\bibinfo {volume} {52}},\ \bibinfo {pages} {1} (\bibinfo {year}
  {1984})%
  \bibAnnoteFile{NoStop}{Bohigas1984}%
\bibitem{Note1}%
  \BibitemOpen
  \bibinfo {note} {The BGS conjecture contains two concepts which deserve
  further explanation. {\protect \emph {K-systems:}} In the mid-1950s
  Kolmogorov introduced a new measure, the metric entropy or K-entropy, to
  classify chaotic systems. Systems with positive K-entropy, such as Arnold's
  famous cat-map, have finite phase space with exponentially separating orbits.
  These systems are called K-systems. {\protect \emph {GOE:}} The central
  objects of random matrix theory are the special ensembles of matrices. These
  ensembles are characterized by the size, the symmetry of the matrices and the
  domain of the individual random elements of these matrices, e.g. GOE is an
  abbreviation for the Gaussian Orthogonal Ensemble consisting real symmetric
  matrices for which the correlated probability of the diagonal elements is
  invariant under orthogonal transformation and follow a multi-dimensional
  normal distribution. This ensemble may, for example, correspond to spinless
  systems, invariant under time reversal.}%
  \bibAnnoteFile{Stop}{Note1}%
\bibitem{osszefoglalo}%
  \BibitemOpen
  \bibfield{author}{%
  in\ \emph{\bibinfo {booktitle} {Proceedings of the Les Houches Summer School
  of Theoretical Physics}},\ }%
  \bibinfo {editor} {edited by\ \bibinfo {editor} {\bibfnamefont{M.~J.}\
  \bibnamefont{Giannoni}}, \bibinfo {editor}
  {\bibfnamefont{A.}~\bibnamefont{Voros}},\ and\ \bibinfo {editor}
  {\bibfnamefont{J.}~\bibnamefont{Zinn-Justin}}}\ (\bibinfo {publisher}
  {(Elsevier, New York, 1991)},\ \bibinfo {year} {1989})%
  \bibAnnoteFile{NoStop}{osszefoglalo}%
\bibitem{HannayAlmeida1984}%
  \BibitemOpen
  \bibfield{author}{%
  \bibinfo {author} {\bibfnamefont{J.}~\bibnamefont{Hannay}}\ and\ \bibinfo
  {author} {\bibfnamefont{A.~M.~O.}\ \bibnamefont{de~Almeida}},\ }%
  \bibfield{journal}{%
  \bibinfo {journal} {J. Phys. A}\ }%
  \textbf{\bibinfo {volume} {17}},\ \bibinfo {pages} {3429} (\bibinfo {year}
  {1984})%
  \bibAnnoteFile{NoStop}{HannayAlmeida1984}%
\bibitem{Berry1985}%
  \BibitemOpen
  \bibfield{author}{%
  \bibinfo {author} {\bibfnamefont{M.~V.}\ \bibnamefont{Berry}},\ }%
  \bibfield{journal}{%
  \bibinfo {journal} {Proc. Roy. Soc. Lond. A}\ }%
  \textbf{\bibinfo {volume} {400}},\ \bibinfo {pages} {229} (\bibinfo {year}
  {1985})%
  \bibAnnoteFile{NoStop}{Berry1985}%
\bibitem{Andreev1996a}%
  \BibitemOpen
  \bibfield{author}{%
  \bibinfo {author} {\bibfnamefont{A.~V.}\ \bibnamefont{Andreev}}, \bibinfo
  {author} {\bibfnamefont{O.}~\bibnamefont{Agam}}, \bibinfo {author}
  {\bibfnamefont{B.~D.}\ \bibnamefont{Simons}},\ and\ \bibinfo {author}
  {\bibfnamefont{B.~L.}\ \bibnamefont{Altshuler}},\ }%
  \bibfield{journal}{%
  \Doi{10.1103/PhysRevLett.76.3947}{\bibinfo {journal} {Phys. Rev. Lett.}}\ }%
  \textbf{\bibinfo {volume} {76}},\ \bibinfo {pages} {3947} (\bibinfo {year}
  {1996})%
  \bibAnnoteFile{NoStop}{Andreev1996a}%
\bibitem{Andreev1996b}%
  \BibitemOpen
  \bibfield{author}{%
  \bibinfo {author} {\bibfnamefont{A.~V.}\ \bibnamefont{Andreev}}, \bibinfo
  {author} {\bibfnamefont{B.~D.}\ \bibnamefont{Simons}}, \bibinfo {author}
  {\bibfnamefont{O.}~\bibnamefont{Agam}},\ and\ \bibinfo {author}
  {\bibfnamefont{B.~L.}\ \bibnamefont{Altshuler}},\ }%
  \bibfield{journal}{%
  \Doi{doi:10.1016/S0550-3213(96)00473-7}{\bibinfo {journal} {Nuclear Physics
  B}}\ }%
  \textbf{\bibinfo {volume} {482}},\ \bibinfo {pages} {536} (\bibinfo {year}
  {1996})%
  \bibAnnoteFile{NoStop}{Andreev1996b}%
\bibitem{Muller2004}%
  \BibitemOpen
  \bibfield{author}{%
  \bibinfo {author} {\bibfnamefont{S.}~\bibnamefont{M\"uller}}, \bibinfo
  {author} {\bibfnamefont{S.}~\bibnamefont{Heusler}}, \bibinfo {author}
  {\bibfnamefont{P.}~\bibnamefont{Braun}}, \bibinfo {author}
  {\bibfnamefont{F.}~\bibnamefont{Haake}},\ and\ \bibinfo {author}
  {\bibfnamefont{A.}~\bibnamefont{Altland}},\ }%
  \bibfield{journal}{%
  \Doi{10.1103/PhysRevLett.93.014103}{\bibinfo {journal} {Phys. Rev. Lett.}}\
  }%
  \textbf{\bibinfo {volume} {93}},\ \bibinfo {pages} {014103} (\bibinfo {year}
  {2004})%
  \bibAnnoteFile{NoStop}{Muller2004}%
\bibitem{Muller2005}%
  \BibitemOpen
  \bibfield{author}{%
  \bibinfo {author} {\bibfnamefont{S.}~\bibnamefont{M{\"u}ller}}, \bibinfo
  {author} {\bibfnamefont{S.}~\bibnamefont{Heusler}}, \bibinfo {author}
  {\bibfnamefont{P.}~\bibnamefont{Braun}}, \bibinfo {author}
  {\bibfnamefont{F.}~\bibnamefont{Haake}},\ and\ \bibinfo {author}
  {\bibfnamefont{A.}~\bibnamefont{Altland}},\ }%
  \bibfield{journal}{%
  \Doi{10.1103/PhysRevE.72.046207}{\bibinfo {journal} {Phys. Rev. E}}\ }%
  \textbf{\bibinfo {volume} {72}},\ \bibinfo {pages} {046207} (\bibinfo {year}
  {2005})%
  \bibAnnoteFile{NoStop}{Muller2005}%
\bibitem{Keating2007}%
  \BibitemOpen
  \bibfield{author}{%
  \bibinfo {author} {\bibfnamefont{J.~P.}\ \bibnamefont{Keating}}\ and\
  \bibinfo {author} {\bibfnamefont{S.}~\bibnamefont{M\"uller}},\ }%
  \bibfield{journal}{%
  \bibinfo {journal} {Proc. Roy. Soc. Lond. A}\ }%
  \textbf{\bibinfo {volume} {463}},\ \bibinfo {pages} {3241} (\bibinfo {year}
  {2007})%
  \bibAnnoteFile{NoStop}{Keating2007}%
\bibitem{kivetel}%
  \BibitemOpen
  \bibfield{author}{%
  \bibinfo {author} {\bibfnamefont{E.~B.}\ \bibnamefont{Bogomolny}}, \bibinfo
  {author} {\bibfnamefont{B.}~\bibnamefont{Georgeot}}, \bibinfo {author}
  {\bibfnamefont{M.-J.}\ \bibnamefont{Giannoni}},\ and\ \bibinfo {author}
  {\bibfnamefont{C.}~\bibnamefont{Schmit}},\ }%
  \bibfield{journal}{%
  \bibinfo {journal} {Phys. Rep.}\ }%
  \textbf{\bibinfo {volume} {291}},\ \bibinfo {pages} {219} (\bibinfo {year}
  {1997})%
  \bibAnnoteFile{NoStop}{kivetel}%
\bibitem{Wu1990}%
  \BibitemOpen
  \bibfield{author}{%
  \bibinfo {author} {\bibfnamefont{H.}~\bibnamefont{Wu}}, \bibinfo {author}
  {\bibfnamefont{M.}~\bibnamefont{Valli{\`e}res}}, \bibinfo {author}
  {\bibfnamefont{D.~H.}\ \bibnamefont{Feng}},\ and\ \bibinfo {author}
  {\bibfnamefont{D.~W.~L.}\ \bibnamefont{Sprung}},\ }%
  \bibfield{journal}{%
  \Doi{10.1103/PhysRevA.42.1027}{\bibinfo {journal} {Phys. Rev. A}}\ }%
  \textbf{\bibinfo {volume} {42}},\ \bibinfo {pages} {1027} (\bibinfo {year}
  {1990})%
  \bibAnnoteFile{NoStop}{Wu1990}%
\bibitem{Arnold}%
  \BibitemOpen
  \bibfield{author}{%
  \bibinfo {author} {\bibfnamefont{V.~I.}\ \bibnamefont{Arnold}},\ }%
  \emph{\bibinfo {title} {Mathematical methods of classical mechanics}},\
  \bibinfo {edition} {2nd}\ ed.,\ \bibinfo {series} {Graduate Texts in
  Mathematics}, Vol.~\bibinfo {volume} {60}\ (\bibinfo {publisher} {Springer
  Verlag},\ \bibinfo {address} {New York},\ \bibinfo {year} {1989})%
  \bibAnnoteFile{NoStop}{Arnold}%
\bibitem{Wu93}%
  \BibitemOpen
  \bibfield{author}{%
  \bibinfo {author} {\bibfnamefont{H.}~\bibnamefont{Wu}}\ and\ \bibinfo
  {author} {\bibfnamefont{D.~W.}\ \bibnamefont{Sprung}},\ }%
  \bibfield{journal}{%
  \Doi{10.1103/PhysRevE.48.2595}{\bibinfo {journal} {Phys. Rev. E}}\ }%
  \textbf{\bibinfo {volume} {48}},\ \bibinfo {pages} {2595} (\bibinfo {year}
  {1993})%
  \bibAnnoteFile{NoStop}{Wu93}%
\bibitem{Ramani1995}%
  \BibitemOpen
  \bibfield{author}{%
  \bibinfo {author} {\bibfnamefont{A.}~\bibnamefont{Ramani}}, \bibinfo {author}
  {\bibfnamefont{B.}~\bibnamefont{Grammaticos}},\ and\ \bibinfo {author}
  {\bibfnamefont{E.}~\bibnamefont{Caurier}},\ }%
  \bibfield{journal}{%
  \Doi{10.1103/PhysRevE.51.6323}{\bibinfo {journal} {Phys. Rev. E}}\ }%
  \textbf{\bibinfo {volume} {51}},\ \bibinfo {pages} {6323} (\bibinfo {year}
  {1995})%
  \bibAnnoteFile{NoStop}{Ramani1995}%
\bibitem{Wu95}%
  \BibitemOpen
  \bibfield{author}{%
  \bibinfo {author} {\bibfnamefont{H.}~\bibnamefont{Wu}}\ and\ \bibinfo
  {author} {\bibfnamefont{D.~W.}\ \bibnamefont{Sprung}},\ }%
  \bibfield{journal}{%
  \Doi{10.1103/PhysRevE.51.6327}{\bibinfo {journal} {Phys. Rev. E}}\ }%
  \textbf{\bibinfo {volume} {51}},\ \bibinfo {pages} {6327} (\bibinfo {year}
  {1995})%
  \bibAnnoteFile{NoStop}{Wu95}%
\bibitem{Brandon03}%
  \BibitemOpen
  \bibfield{author}{%
  \bibinfo {author} {\bibfnamefont{B.~P.}\ \bibnamefont{van Zyl}}\ and\
  \bibinfo {author} {\bibfnamefont{D.~A.}\ \bibnamefont{Hutchinson}},\ }%
  \bibfield{journal}{%
  \Doi{10.1103/PhysRevE.67.066211}{\bibinfo {journal} {Physical Review E}}\ }%
  \textbf{\bibinfo {volume} {67}},\ \bibinfo {pages} {066211} (\bibinfo {year}
  {2003})%
  \bibAnnoteFile{NoStop}{Brandon03}%
\bibitem{Schumayer2008}%
  \BibitemOpen
  \bibfield{author}{%
  \bibinfo {author} {\bibfnamefont{D.}~\bibnamefont{Schumayer}}, \bibinfo
  {author} {\bibfnamefont{B.~P.}\ \bibnamefont{van Zyl}},\ and\ \bibinfo
  {author} {\bibfnamefont{D.~A.~W.}\ \bibnamefont{Hutchinson}},\ }%
  \bibfield{journal}{%
  \Doi{10.1103/PhysRevE.78.056215}{\bibinfo {journal} {Phys. Rev. E}}\ }%
  \textbf{\bibinfo {volume} {78}},\ \bibinfo {pages} {056215} (\bibinfo {year}
  {2008})%
  \bibAnnoteFile{NoStop}{Schumayer2008}%
\bibitem{Schumayer2011}%
  \BibitemOpen
  \bibfield{author}{%
  \bibinfo {author} {\bibfnamefont{D.}~\bibnamefont{Schumayer}}\ and\ \bibinfo
  {author} {\bibfnamefont{D.~A.~W.}\ \bibnamefont{Hutchinson}},\ }%
  \bibfield{journal}{%
  \Doi{10.1103/RevModPhys.83.307}{\bibinfo {journal} {Rev. Mod. Phys.}}\ }%
  \textbf{\bibinfo {volume} {83}},\ \bibinfo {pages} {307} (\bibinfo {year}
  {2011})%
  \bibAnnoteFile{NoStop}{Schumayer2011}%
\bibitem{Note2}%
  \BibitemOpen
  \bibinfo {note} {Unfolding is a standard procedure in RMT to scale the mean
  level spacing of a given spectrum to be unity, thus different spectra become
  comparable.}%
  \bibAnnoteFile{Stop}{Note2}%
\bibitem{Shabat1992}%
  \BibitemOpen
  \bibfield{author}{%
  \bibinfo {author} {\bibfnamefont{A.}~\bibnamefont{Shabat}},\ }%
  \bibfield{journal}{%
  \Doi{10.1088/0266-5611/8/2/009}{\bibinfo {journal} {Inverse Problems}}\ }%
  \textbf{\bibinfo {volume} {8}},\ \bibinfo {pages} {303} (\bibinfo {year}
  {1992})%
  \bibAnnoteFile{NoStop}{Shabat1992}%
\bibitem{Borg1946}%
  \BibitemOpen
  \bibfield{author}{%
  \bibinfo {author} {\bibfnamefont{G.}~\bibnamefont{Borg}},\ }%
  \bibfield{journal}{%
  \Doi{10.1007/BF02421600}{\bibinfo {journal} {Acta Mathematica}}\ }%
  \textbf{\bibinfo {volume} {78}},\ \bibinfo {pages} {1} (\bibinfo {year}
  {1946})%
  \bibAnnoteFile{NoStop}{Borg1946}%
\bibitem{Zakhariev1990}%
  \BibitemOpen
  \bibfield{author}{%
  \bibinfo {author} {\bibfnamefont{B.~N.}\ \bibnamefont{Zakhariev}}\ and\
  \bibinfo {author} {\bibfnamefont{A.~A.}\ \bibnamefont{Suzko}},\ }%
  \emph{\bibinfo {title} {{D}irect and {I}nverse {P}roblems: {P}otentials in
  {Q}uantum {S}cattering}}\ (\bibinfo {publisher} {Springer},\ \bibinfo {year}
  {1990})%
  \bibAnnoteFile{NoStop}{Zakhariev1990}%
\bibitem{Harper1955}%
  \BibitemOpen
  \bibfield{author}{%
  \bibinfo {author} {\bibfnamefont{P.~G.}\ \bibnamefont{Harper}},\ }%
  \bibfield{journal}{%
  \Doi{http://dx.doi.org/10.1088/0370-1298/68/10/304}{\bibinfo {journal} {Proc.
  Phys. Soc. A}}\ }%
  \textbf{\bibinfo {volume} {68}},\ \bibinfo {pages} {874} (\bibinfo {year}
  {1955})%
  \bibAnnoteFile{NoStop}{Harper1955}%
\bibitem{Hofstadter1976}%
  \BibitemOpen
  \bibfield{author}{%
  \bibinfo {author} {\bibfnamefont{D.~R.}\ \bibnamefont{Hofstadter}},\ }%
  \bibfield{journal}{%
  \Doi{10.1103/PhysRevB.14.2239}{\bibinfo {journal} {Phys. Rev. B}}\ }%
  \textbf{\bibinfo {volume} {14}},\ \bibinfo {pages} {2239} (\bibinfo {year}
  {1976})%
  \bibAnnoteFile{NoStop}{Hofstadter1976}%
\end{thebibliography}%

\end{document}